\documentclass[12pt,preprint]{aastex}
\usepackage[usenames,dvips]{color}
\usepackage{natbib}
\slugcomment{}
\shorttitle{Hydrodynamical Simulations of NGC 1097 
} \shortauthors{Lien-Hsuan Lin et al.}

\begin{document}

\title{Hydrodynamical Simulations of the Barred Spiral Galaxy NGC 1097}
\author{Lien-Hsuan Lin\altaffilmark{1}, Hsiang-Hsu Wang\altaffilmark{1}, Pei-Ying Hsieh\altaffilmark{1,2}, Ronald E. Taam\altaffilmark{1,3}, Chao-Chin Yang\altaffilmark{4}, and David C. C. Yen\altaffilmark{5}}
\altaffiltext{1}{Institute of Astronomy and Astrophysics, Academia Sinica, P.O. Box 23-141, Taipei 10617, Taiwan, R.O.C.}
\altaffiltext{2}{Institute of Astrophysics, National Central University, Jhongli City, Taoyuan County 32001, Taiwan, R. O. C.}
\altaffiltext{3}{Department of Physics and Astronomy, Northwestern University, 2131 Tech Drive, Evanston, IL 60208, USA}
\altaffiltext{4}{Lund Observatory, Department of Astronomy and Theoretical Physics, Lund University, Box 43, Lund 221 00, Sweden}
\altaffiltext{5}{Department of Mathematics, Fu Jen Catholic University, Taipei 24205, Taiwan, R.O.C.}

\begin{abstract}
NGC 1097 is a nearby barred spiral galaxy believed to be interacting with the elliptical galaxy NGC 1097A located to its northwest.
 It hosts a Seyfert 1 nucleus surrounded by a circumnuclear starburst ring.
 Two straight dust lanes connected to the ring extend almost continuously out to the bar. 
The other ends of the dust lanes attach to two main spiral arms.
 To provide a physical understanding of its structural and kinematical properties, 
two-dimensional hydrodynamical simulations have been carried out. 
Numerical calculations reveal that many features of the gas morphology and kinematics can be reproduced 
provided that the gas flow is governed by a gravitational potential associated with a slowly rotating  strong bar. 
By including the self-gravity of the gas disk in our calculation, we have found the starburst ring to be gravitationally unstable which is consistent with the observation in \citet{hsieh11}. 
Our simulations show that the gas inflow rate is 0.17 M$_\sun$ yr$^{-1}$ 
into the region within the starburst ring even after its formation, leading to the coexistence of both a nuclear 
ring and a circumnuclear disk.
\end{abstract}
\keywords{galaxies: individual(NGC 1097) --- galaxies: kinematics and dynamics --- galaxies: spiral --- galaxies: structure --- galaxies: evolution --- galaxies: Seyfert --- galaxies: starburst}

\section{Introduction}
It has long been recognized that a periodic potential, such as a rotating bar, can drive density waves in a differentially rotating disk and transport gas toward galactic centers \citep[e.g.,][]{athan92, piner95, yuankuo97, macie04}.
As pointed out by \citet{yuankuo97}, density waves excited by the bar at the outer inner Lindblad resonance (OILR) and the outer Lindblad resonance (OLR) will propagate inward and outward, respectively.
The waves excited at the OILR carry negative angular momentum which will be deposited over the annular region covered by the waves during propagation.
The disk material there, after losing angular momentum, would flow toward the center.
The bar-driven gas transported into the central regions of galaxies can fuel nuclear star formation and active galatic nuclei (AGNs).
Indeed, statistical evidence based on the observations of a number of galaxies shows that 
barred galaxies have a higher fraction
of central concentration of molecular gas in their centers than non-barred galaxies \citep{sakamoto99}.
Their result supports the theory of bar-driven gas transport mentioned above.

Subsequent observations at higher angular resolution were able to resolve the central molecular concentration of galaxies into a wide variety of morphologies.
The NUclei of GAlaxies (NUGA) survey revealed, for example, that 
NGC 7217 has a nuclear ring of 1.8 kpc in diameter and a nuclear spiral \citep{boone04}, NGC 4826 has two one-arm trailing spirals at different radii and a lopsided nuclear disk of 40 pc radius \citep{boone04}, NGC 4579 has two leading gas lanes inside 500 pc and a 150 pc off-centered ringed disk \citep{garcia04}, NGC 6951 has two nuclear spiral arms in the inner 700 pc region \citep{garcia04}, NGC 4321 has two nuclear spiral arms connected to a 150 pc disk \citep{garcia05}, while NGC 6574 has a nuclear gas bar and spirals \citep{lindt08}.  
Many nuclear rings exhibit regions of 
intense star formation \citep[e.g.,][]{maoz96}, which suggests that the rings are 
characterized by sufficiently high surface densities resulting in gravitational instabilities.
It is generally believed that such starburst rings are a consequence of the accumulation 
of inflowing gas and dust driven by a non-axisymmetric potential from a stellar bar.

NGC 1097 is one of the most well-studied galaxies containing a central starburst ring 
\citep{hummel87, telesco93, barth95, quillen95, storchi96, kotilainen00, sandstrom10}.
Inside the ring, a relatively strong molecular concentration is detected in $^{12}$CO (2-1) emission by \citet{hsieh08} using  the Submillimeter Array \footnote{The SMA is a joint project between the Smithsonian Astrophysical Observatory and the Academia Sinica Institute of Astronomy and Astrophysics and is funded by the Smithsonian Institution and the Academia Sinica.}\citep{ho04}. 
Hence, NGC 1097 is a good candidate to test the theory of bar-driven gas transport.
However, the coexistence of the nuclear ring and the circumnuclear disk as observed in NGC 1097 has not yet been 
seen in previous hydrodynamical simulations of bar galaxies \citep[e.g.,][]{athan92, wada92, piner95, patsis00, macie04}.
In this study, we apply our hydrodynamical model with the self-gravity of the gas to simulate NGC 1097, reproducing 
its circumnuclear disk, nuclear ring as well as other observed features.

In the next section, the observational data of NGC 1097 are summarized. The numerical method for the hydrodynamical simulations are described in \S 3. 
The procedures adopted for the theoretical modeling of the observational data are outlined in \S 4,
  and the comparisons of the numerical results with the data are presented in \S 5.
Finally, we conclude in \S 6.

\section{Observational data}
\begin{figure}
\figurenum{1}
\epsscale{0.7}
\plotone{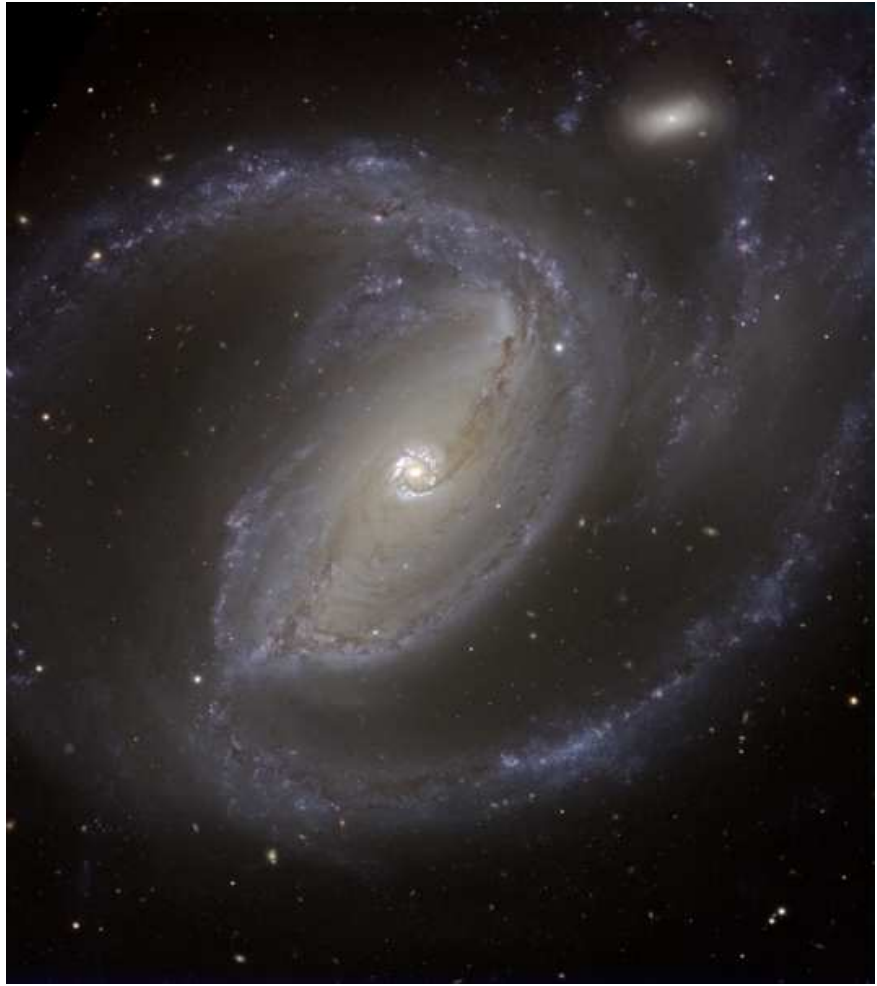}
\caption{Optical image of NGC 1097 taken with the Visible Multi-Object Spectrograph (VIMOS) instrument on 
the 8.2-m Melipal (Unit Telescope 3) of ESO's Very Large Telescope. North is at the top and east is to 
the left.  NGC 1097 exhibits a bright nucleus associated with an AGN and a nuclear starburst ring 
consisting of bright knots at a radius of 10$\arcsec$ ($\sim$ 820 pc).  These knots are usually HII regions 
under intense radiation from luminous massive OB stars.  The presence of these knots suggests a vigorous 
burst of star formation has occurred recently.  NGC 1097 is believed to be interacting with the elliptical 
galaxy NGC 1097A located to its north-west.}
\label{Oimage1097}
\end{figure}

NGC 1097 is a nearby barred spiral galaxy inclined at an angle of 46$^{\circ}$ \citep{ondrechen89}.  
With a redshift of $0.00424\pm0.00001$ \citep{koribalski04} and Hubble constant of 75 km s$^{-1}$ Mpc$^{-1}$, 
we adopt a distance of 16.96 Mpc, yielding a scale of 1$\arcsec$ = 82.2 pc.
The galaxy is of type SB(s)b \citep{vaucouleurs91} or (R)SB(rs)bc \citep{sandage81}.
Figure~\ref{Oimage1097} shows its optical image taken by the Visible Multi-Object Spectrograph (VIMOS) 
instrument on the 8.2-m Melipal (Unit Telescope 3) of ESO's Very Large Telescope.  The image reveals a 
circumnuclear starburst ring, with a radius of 10$\arcsec$ (0.82 kpc), connected with a pair of nearly 
straight dust lanes.  The other ends of the dust lanes are attached to two spiral arms that can be traced 
continuously for more than 270$^{\circ}$ in azimuth.  The inner parts of these two spirals trace the boundary of the 
bar region and form an oval ring-like structure (inner ring).  The nucleus of NGC 1097 was originally 
identified as a LINER \citep{keel83}, however, over the past two decades, it has exhibited Seyfert 1 activities 
as evidenced by the presence of broad double-peaked Balmer emission lines \citep{storchi93}.
The starburst ring is luminous at wavelengths including radio \citep{hummel87}, mid-infrared \citep{telesco93}, optical \citep{barth95, quillen95, storchi96}, near-infrared \citep{kotilainen00}, and soft X-rays \citep{perez96}.  It is rich in molecular gas (5.8$\times$ 10$^8$ M$_{\odot}$; \citet{hsieh08}) and has a high star formation rate (3.1 M$_{\odot}$ yr$^{-1}$; \citet{hsieh11}).

The $^{12}$CO (J = 2-1) integrated intensity map of the central region of NGC 1097 \citep{hsieh11} is 
shown in Figure~\ref{PYCOMT0}.  A relatively strong molecular concentration (radius $\sim$ 300 pc) is detected at the very center 
surrounded by a weaker molecular ring (radius $\sim$ 820 pc), which spatially coincides with the AGN and the starburst ring 
respectively.  The molecular ring is found to be composed of several knots.  The two strongest emission 
knots are located at the NE and SW sides of the ring in the regions where the dust lanes connect with 
the ring.  Such molecular gas peaks are often seen in barred spiral galaxies, which are known as the 
twin-peak morphology \citep[e.g,][]{kenney92}.

Figure~\ref{HIvf} shows the velocity field of the HI observation by \citet{ondrechen89}.  The most striking 
characteristic in the observation is a series of regular bendings (or wiggles) along the SW arm in the 
isovelocity contours of the velocity field, which provides evidence for the presence of noncircular 
gaseous motions. 
The same characteristic is also present in \citet{higdon03}. However, since the beam size in \citet{higdon03} is twice of that in \citet{ondrechen89}, we have adopted the data observed by \citet{ondrechen89}.
The position angle of the line of nodes derived from the HI observation is found 
to be 134$^{\circ} \pm$ 3$^{\circ}$, while the average inclination obtained by \citet{ondrechen89} is 
46$^{\circ} \pm$ 5$^{\circ}$.

The intensity-weighted $^{12}$CO (J = 2-1) mean-velocity map \citep{hsieh11} is shown in 
Figure~\ref{COvf}.  Both the central molecular concentration and the molecular ring show an overall 
velocity gradient in the NW to SE direction along the major axis of the large-scale galactic disk (line of nodes), indicating the differential rotation of the gas disk.  
The emission is blueshifted on the northwestern side and redshifted on the southeastern side 
of the center.

The basic parameters for NGC 1097 are summarized in Table 1.

\section{Numerical Method}
The numerical method for the hydrodynamical simulations used in our study of NGC 1097 is nearly identical as that used in \citet{lin11}.  
The simulations were performed with a high-order Godunov code known as Antares \citep{yuan05}.
We only briefly point out here the most notable features of the method.
The total gravitational potential in the equation of motion is composed of three components:
\begin{equation}
V = V_0+V_1+V_g.
\end{equation}
The first term, $V_0$, is a central potential supporting a differentially rotating disk deduced from the observed rotation curve.
The second term, $V_1$, is the dominant second harmonic component of a rotating bar potential 
taken in the form,
\begin{equation}
V_1(R,\phi,t)=\Phi(R)\cos[2(\phi-\Omega_pt)],
\end{equation}
where $\Omega_p$ is the angular speed of the bar.
We employ a simple functional form for the amplitude given as:
\begin{equation}
\Phi(R)=-\Phi_0\frac{R^2}{(A^2_1+R^2)^2},
\end{equation}
where $A_1 \equiv a_1/r_s$ , $R \equiv r/r_s$, and $r_s$ = 1.0 kpc. $A_1$ and $R$ are both dimensionless.
The parameter $a_1$ denotes the radial distance at which the
bar potential is a minimum. 
In order to reduce numerical noise, the amplitude of the bar potential, $\Phi_0$, is increased gradually from zero to the full value during the initial two rotation periods of the bar.
 $V_g$, the last term of $V$, represents the self-gravitational
potential of the gaseous disk.
In order to include the self-gravity of the gas disk in the calculation,
the hydrodynamic code is coupled with a Poisson solver.
A description for the method of solution for the Poisson solver in the Antares code can be found in \citet{yen12}.

The method used in this study differs from that used in \citet{lin11} in only four aspects:

1) The initial surface density of the gas was taken to be uniform in \citet{lin11}, while in this work it is 
assumed to follow an exponential law of the form
\begin{equation}
\sigma=\sigma_0e^{-(\frac{r}{r_0})^2},
\end{equation}
where $\sigma_0$ is the initial surface density at the center, taken to be 15 $M_\odot/pc^{2}$.
The value of $r_0$ is 14 kpc.  The total gas mass inside 20 kpc in our simulations is 8.3 $\times$ 
10$^{9}$ M$_\sun$. For comparison, the total molecular mass of NGC 1097 is 9.4 $\times$ 10$^{9}$ M$_\sun$ \citep{crosthwaite02}.	 

2) The initial settings for the three parameters, $v_0, B,$ and $A$, used in the Elmegreen rotation curve 
\citep{elmegreen90} given by 
\begin{equation}
v(r)=v_0(\frac{r}{r^B+r^{1-A}})
\end{equation}
are found to be 550 km s$^{-1}$, 0.552, and $-0.249$ respectively by means of a least-square fit to the 
observational data points of NGC 1097.  The rotation curve and the data points are shown in Figure~
\ref{rotc1097} and the corresponding angular speed curves are shown in Figure~\ref{anspd1097}. 
The horizontal line represents the pattern speed of the bar of our choice, $\Omega_p$ (see Section 4).  The intersections of $\Omega_p$ with 
the $\Omega\pm\kappa/2$ and $\Omega\pm\kappa/4$ curves determine the locations of the outer-inner Lindblad 
resonance (OILR) and the outer Lindblad resonance (OLR) as well as those of the inner and outer 4:1 
resonances.  Here, $\kappa$ is the epicyclic frequency.  Since $\Omega$ - $\kappa/2$ increases monotonically 
with decreasing $r$, an inner-inner Lindblad resonance (IILR) does not exist.

3) The sound speed, $a$, for the gaseous disk is adopted to be equal to 15 km s$^{-1}$. 
The gas response is not very sensitive to this parameter from 10 to 15 km 
s$^{-1}$, both in our tests and in Patsis \& Athanassoula (2000).

4) All the calculations presented in this paper are performed on a 1536 $\times$ 1536 uniform Cartesian grid.  The 
computational domain corresponds to a physical region of 60 kpc $\times$ 60 kpc, resulting in a cell size
of 39 $\times$ 39 pc$^{2}$ (or 0.476$\arcsec \times$0.476$\arcsec$).

\section{Modeling Procedure}
\subsection{Requirements for a Successful Model}
NGC 1097 is clearly not bisymmetrical.  It has an elliptical companion galaxy, NGC 1097A.  
Based on their redshift data, the distance between these two galaxies is estimated to be $\sim$ 1 Mpc.
For both visual and near-infrared wavelengths, the luminosity ratio of NGC 1097A to NGC 1097 is about 0.01. 
From the optical image in Figure~\ref{Oimage1097}, it can be seen that the northern half of NGC 1097 is influenced by NGC 
1097A.   However, the HI velocity fields in Figure~\ref{HIvf} are quite regular, apart from indications of a 
slight asymmetry, which is consistent with a weak tidal interaction.  Therefore, the simulation results 
from our bisymmetrical model can still provide useful information for the morphology and 
kinematics of NGC 1097.  The selection criteria of a successful model are based on the following 
requirements, which are similar to those employed by \citet{lindblad96} and \citet{lindblad96b}:
 
1. The nuclear starburst ring should be reproduced.

2. The gas lanes in the model should have approximately the same shape and position as the 
observed dust lanes.
  
3. The two main spiral arms should be reproduced.

4. The characteristics in the observed velocity field should be reproduced.

The simulation results are judged by visual inspection of the model density distribution and velocity maps 
overlaid on the observed ones.  Such comparisons would be sufficient in the context of our simple model.
A quantitative method, e.g. $\chi^{2}$, can be misleading if the strong density features or steep velocity 
gradients present in the models do not exactly overlap their observed counterparts.  Furthermore, since 
the main purpose of this study is to obtain an understanding of NGC 1097 by simulating as much as possible the observed features using a simple yet physically plausible model,
 it is not our goal in this study to simulate complex and asymmetric features in 
the observations.

\subsection{Free Parameters}

The free parameters underlying our hydrodynamic model which are important for the response of gas to a 
rotating bar are the angular pattern speed, the location of the potential minimum, and the strength of 
the bar potential.  These parameters are varied in our simulations to seek the best-fit models.

The pattern speed $\Omega_p$ controls the positions of the resonances, which in turn determine the morphology of the spiral structure \citep[e.g.,][]{yuankuo97}.  For an initial estimate of the pattern speed, we choose a value that places the inner 4:1 
ultraharmonic resonance (UHR) near the ends of the inner ring and the OILR outside the circumnuclear 
starburst ring.  As pointed out by \citet{schwarz84}, the inner rings are close to the inner ultraharmonic 
resonance where ($\Omega$ - $\kappa$/4) = $\Omega_p$.  This feature is also confirmed by \citet{lin08} 
in their study of NGC 6782.  Furthermore, it has been shown in hydrodynamical simulations that nuclear 
spirals or nuclear rings are formed inside the OILR \citep{athan92, piner95, macie04}.
An analysis of the central region in M100 (NGC 4321) by \citet{knapen95} also shows that the nuclear 
ring should be located inside the OILR.  The range for the pattern speed $\Omega_p$ is thus chosen to be 
20 $\leq \Omega_p \leq$ 22.1 km s$^{-1}$ kpc$^{-1}$, corresponding to a range for the radius of the 
OILR given as 4.0 kpc $\leq$ R$_{OILR}$ $\leq$ 4.44 kpc.

The second free parameter in our model is the location of the minimum of the bar potential (a$_1$ in Eq. 3). 
It will affect the background central mass concentration and the size of the nuclear ring.
A smaller value of a$_{1}$ leads to a smaller  nuclear ring.
In our simulations, a$_1$ is taken to lie in the range 1.9 $\leq$ a$_1 \leq$ 2.3 kpc.  

The third parameter is the strength of the bar potential. 
It is represented by $f_{OILR}$, which is defined as the ratio of the effective radial force 
exerted on the disk by the rotating bar \citep{yuankuo97} to the force required for circular motion as inferred from 
the rotation curve at the location of OILR.  This choice is motivated by the fact that the waves are 
excited at the resonances.  
Its main influence is on the shape of the gas lanes. 
A stronger bar will produce straighter gas lanes. 
A similar effect is shown by the quadrupole moment in \citet{athan92}.
We found in our simulations that $f_{OILR}$ should lie between 16\% and 23\%.

\section{Results}
\subsection{Evolution of the gaseous disk}
To understand the formation of the steady state structure, the 
sequential images of the evolution of the density features are shown in Figure~\ref{evolution1097}.
Initially, spiral waves excited at the OILR are the first to appear (Figure~\ref{evolution1097}b), and
as the strength of the bar potential increases,  these waves intensify (Figure~\ref{evolution1097}c), 
eventually leading to the formation of a shock (Figure~\ref{evolution1097}d). 
Subsequently, the inner parts of the spirals excited at the OILR become 
tightly wound to form an elliptical ring, which we refer to be the circumnuclear ring (Figure~\ref{evolution1097}e).  
At the same time, the waves excited at the inner 4:1 resonance appear 
(the two patches located on the 4:1 UHR circle in the first and third quadrants at 45 degrees in Figure~\ref{evolution1097}e), 
grow in strength, and propagate inward.
Thereafter, they interact with the 
waves excited at the OILR and merge at the gas lanes (Figure~\ref{evolution1097}f).  
In the meantime, the waves excited at the OLR also appear.
As a result of the self-gravity included in our simulation, the circumnuclear ring becomes unstable when the surface density is sufficiently high so that the Toomre Q-value is less than one (see Section 5.3).  
  After two and a half revolutions of the bar, corresponding to an evolution time of 711 Myr (Figure~\ref{evolution1097}g), the 
morphology of the gaseous disk remains nearly unchanged except for a small periodic change in the morphology 
of the circumnuclear ring due to the nonlinear development of the gravitational instability.

\subsection{Comparisons between numerical results and observations}

Based on 25 simulations, we deduced that the values of the parameters which provide 
the best fit to the observations are $\Omega_p$ = 21.6 km s$^{-1}$ kpc$^{-1}$, a$_1$ = 2.2 kpc, and $f_{OILR}$ 
= 21\%.  The value for $\Omega_p$ implies that the OILR, the UHR, the CR (corotation resonance), and the OLR are located 
at R$_{OILR}$ = 4.09 kpc, R$_{UHR}$ = 8.17 kpc, R$_{CR}$ = 11.7 kpc, and R$_{OLR}$ = 18.0 kpc respectively. 

For comparison to the observed data, we choose the frame corresponding to a time after 4.35 revolutions of the 
bar, or 1237 Myr (Figure~\ref{evolution1097}h).  For ease of presentation, 
we orient the simulation results such that
the P.A. of the line of nodes, the inclination, and the P.A. of the major bar of the galaxy are 134$^{\circ}$, 46$^{\circ}$, and 151$^{\circ}$ respectively as listed in Table 1.

\subsubsection{Images}
The projected surface density distribution for the best-fit case is illustrated in the left panel of 
Figure~\ref{surfden1097} and superimposed onto Figure~\ref{Oimage1097} (reproduced in the middle panel) 
in the right panel of Figure~\ref{surfden1097}. It can be seen that the circumnuclear starburst ring, 
the northwest dust lane, the two lateral spiral arms along the bar region as well as the southwest main spiral arm 
are reproduced by the simulation in both their shapes and locations.  The circumnuclear ring is reproduced by 
the tightly wound inner parts of the spirals excited at the OILR and is located inside the OILR as expected \citep{athan92, piner95, macie04, yuan06}.
The lateral arms along the bar region correspond to the 4:1 ultraharmonic spirals in our simulation.
The lateral spiral arm on the southwestern side in the observation reveals that dust is distributed on the 
concave side of the spiral and that the bright OB stars lie outside the dust.  This is consistent with the 
theory \citep[e.g.,][]{roberts69} that inside the corotation radius dust coincides with the shock on the concave 
side of the spiral arm, gas is compressed by the shock, and after the gas has passed through the shock luminous 
stars are formed.  The shock associated with the 4:1 ultraharmonic arm in our simulation coincides with the 
observed dust and bright OB stars lying behind the shock.  For the main spiral arms outside the bar region in 
the optical image, the southwest spiral arm coincides with the spiral excited at the OLR in our simulation.
However, the spatial extent of the north-east spiral arm is not as large as the arm in the southwest and is 
marked by a discontinuity.  In fact, only a small portion of the observed arm matches with the simulated spiral 
excited at the OLR.  Since the northern half of NGC 1097 is likely influenced by NGC 1097A, the complexity and 
asymmetry of the feature in the observation is difficult to simulate within the framework of our simple model.  

The projected simulated density distribution convolved with the synthesized beam in the HI map produced by 
\citet{ondrechen89} is shown in the left panel of Figure~\ref{surfden1097HI}. Its superposition onto the 
observed HI map (the middle panel of Figure~\ref{surfden1097HI}) is shown in the right panel of the same figure.
The southwest arm in the HI map matches quite well with the simulation result,  which is consistent with the 
comparison between the optical image and the simulation.  The HI intensity peak located at the southern end of 
the bar in the optical image coincides with the local high density region in the simulation.  Although the 
simulated north-east spiral does not match well with the optical spiral arm, it matches much better with the HI 
arm which extends further than the optical arm.  Since HI is usually absent in centers of galaxies, we may 
attribute the central high density region in our simulation to CO gas (although different chemical species 
in the gas are not distinguished in our simulation).

\begin{figure}
\figurenum{2}
\epsscale{0.75}
\plotone{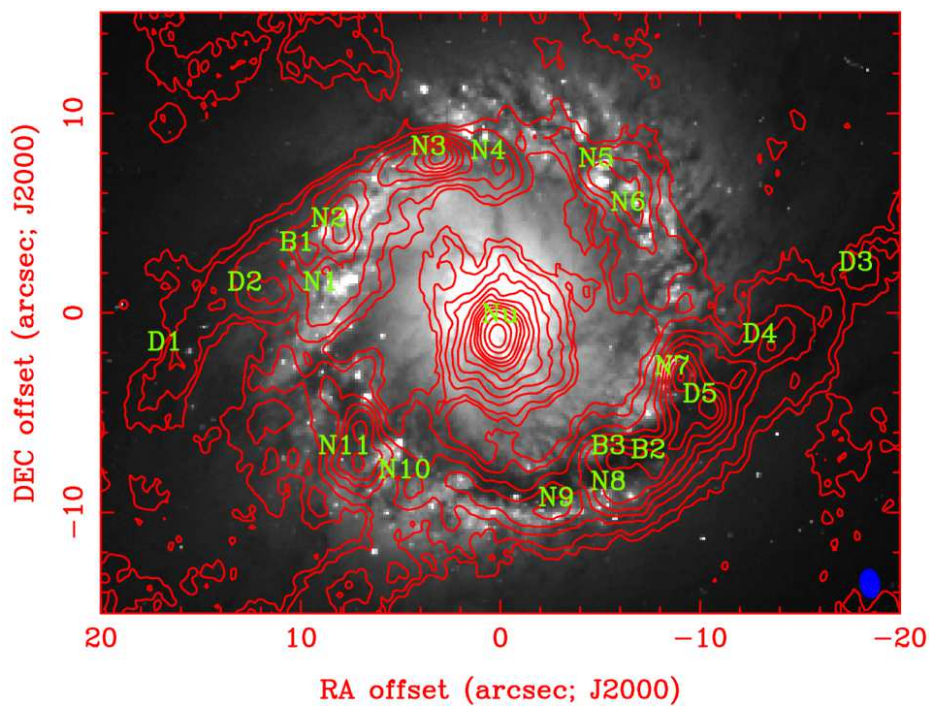}
\caption{Top panel of Figure 4 in \citet{hsieh11}. 
This image is the $^{12}$CO(J = 2-1)-integrated map (contours) overlaid on the archival $HST I$-band (Filter F814 W) image (grayscale). 
The contour levels for $^{12}$CO(J = 2-1) are 2$\sigma$, 3$\sigma$, 5$\sigma$,..., 20$\sigma$, 25$\sigma$, and 30$\sigma$ (1$\sigma$ = 2.3 Jy km s$^{-1}$ beam$^{-1}$).
The IDs for the individual peaks of clumps are marked.
The CO-synthesized beam (1$\arcsec$.5 $\times$ 1$\arcsec$.0, P.A. = 8$^{\circ}$.1) is shown in the lower right corner.
  North is at the top and east is to the left.}
\label{PYCOMT0}
\end{figure}

\begin{figure}
\figurenum{3}
\epsscale{0.6}
\plotone{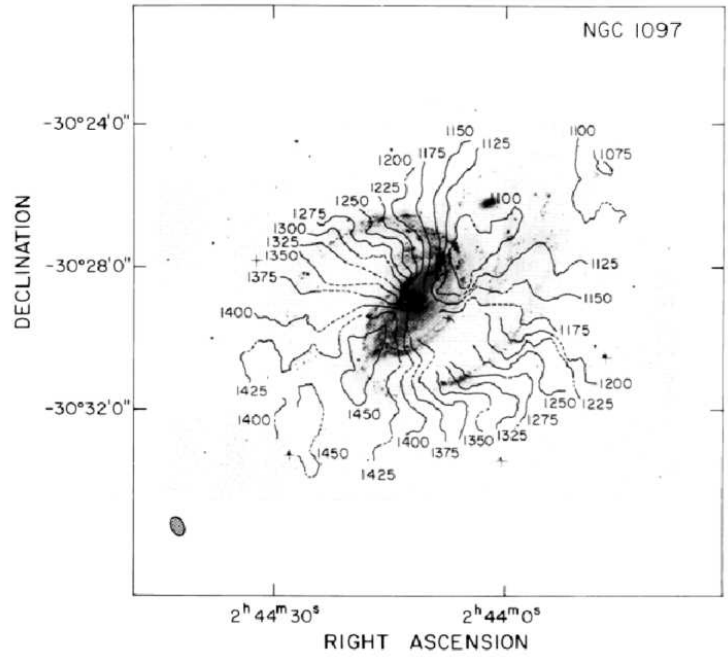}
\caption{Isovelocity contour plot of the velocity field in NGC 1097 superposed on a photograph of the galaxy (Ondrechen et al. 1989).
The wiggles along the spiral arms in the isovelocity contours of the velocity field provide evidence for the presence of noncircular motions of the gas.
Isovelocity contours are labelled in km s$^{-1}$.  North is at the top and east is to the left.}

\label{HIvf}
\end{figure}

\begin{figure}
\figurenum{4}
\epsscale{0.75}
\plotone{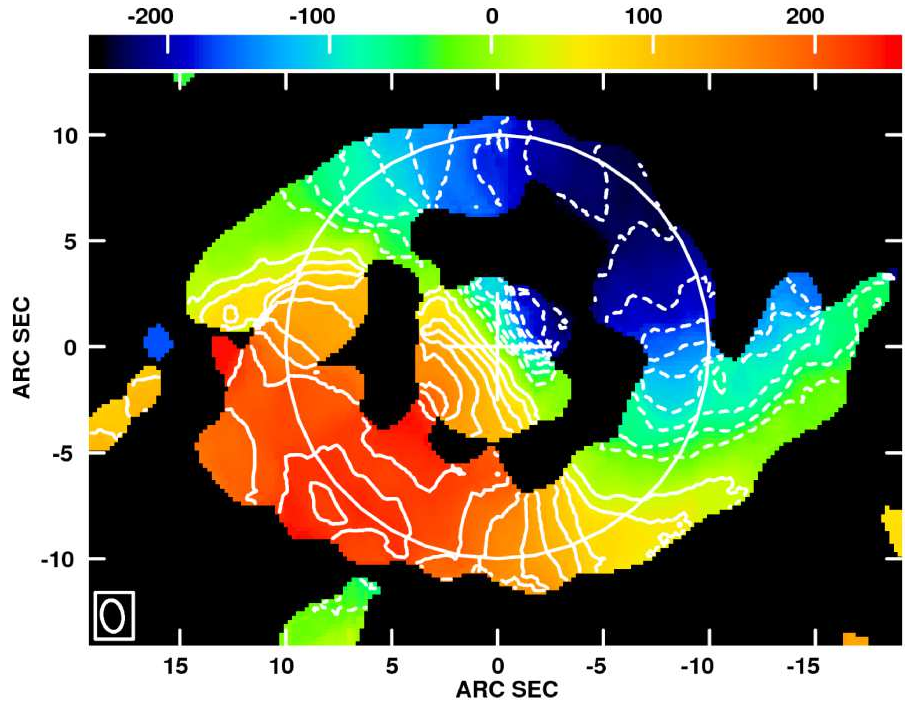}
\caption{
Intensity-weighted mean velocity (moment 1) map of $^{12}$CO(J = 2-1) \citep{hsieh11}. The contour interval is 25 km s$^{-1}$.
The map shows a velocity gradient from redshift to blueshift, as indicated by the color bar.
The nuclear gas is rotating in the same sense with the molecular ring.  North is at the top and east is to the left.}
\label{COvf}
\end{figure}

\begin{figure}
\figurenum{5}
\epsscale{1.0}
\plotone{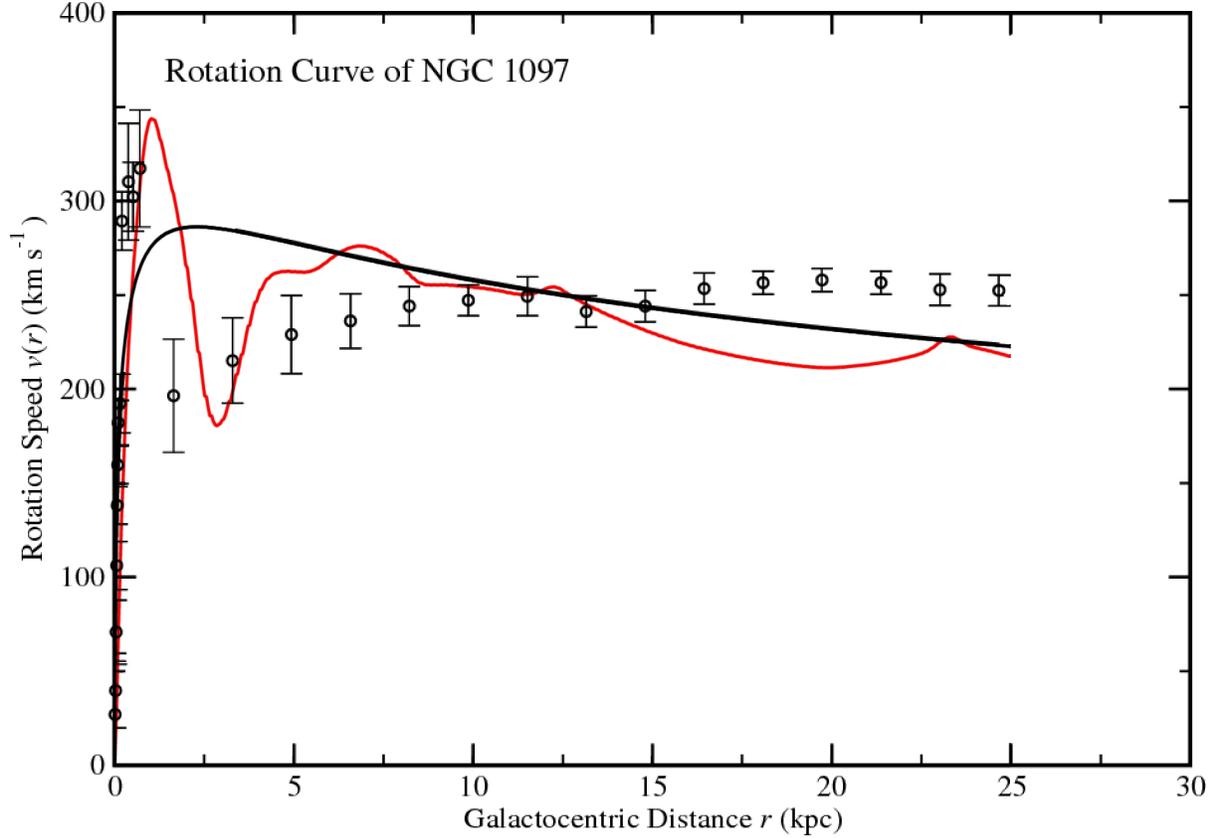}
\caption{Adopted representation of the initial rotation curve of NGC 1097 (black solid line) for the axisymmetric force in our model.
The observational data points (represented by circles in the figure) consist of two groups: those within a distance of 1 kpc are from $^{12}$CO(J=2-1) observations \citep{hsieh08}, while the rest are from HI observations (Ondrechen et al. 1989). The data have been corrected for inclination.
The red line is the evolved rotation curve (convolved with the synthesized beam in $^{12}$CO(J=2-1) observation) along the major axis at the time we select for comparing the simulation result with observations.}
\label{rotc1097}
\end{figure}

\begin{figure}
\figurenum{6}
\epsscale{1.0}
\plotone{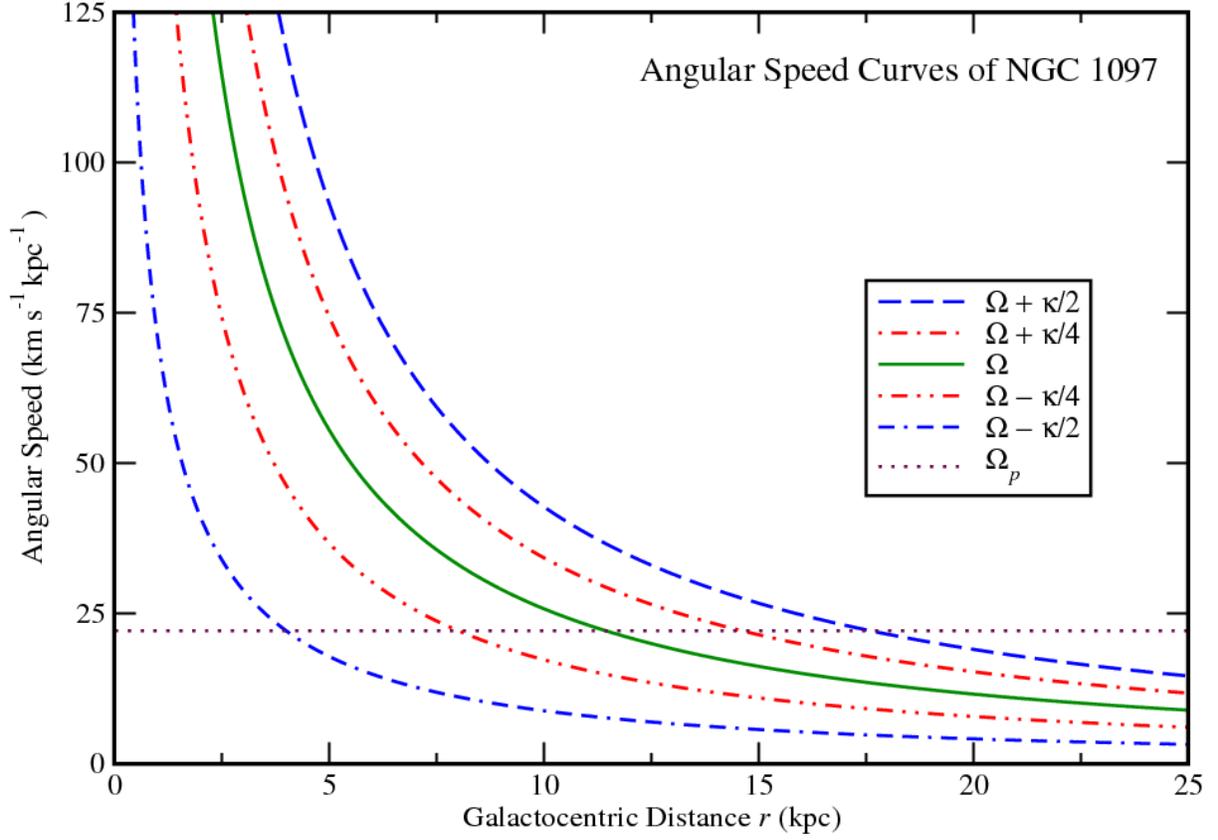}
\caption{Angular speed curves as functions of radius derived from the Elmegreen rotation curve for an angular pattern speed of the bar $\Omega_p$ = 21.6 km s$^{-1}$ kpc$^{-1}$. $\Omega$ and 
$\kappa$ are the circular angular speed and the radial epicyclic frequency, respectively.} 
\label{anspd1097}
\end{figure}

\begin{figure}
\figurenum{7}
\epsscale{0.68}
\plotone{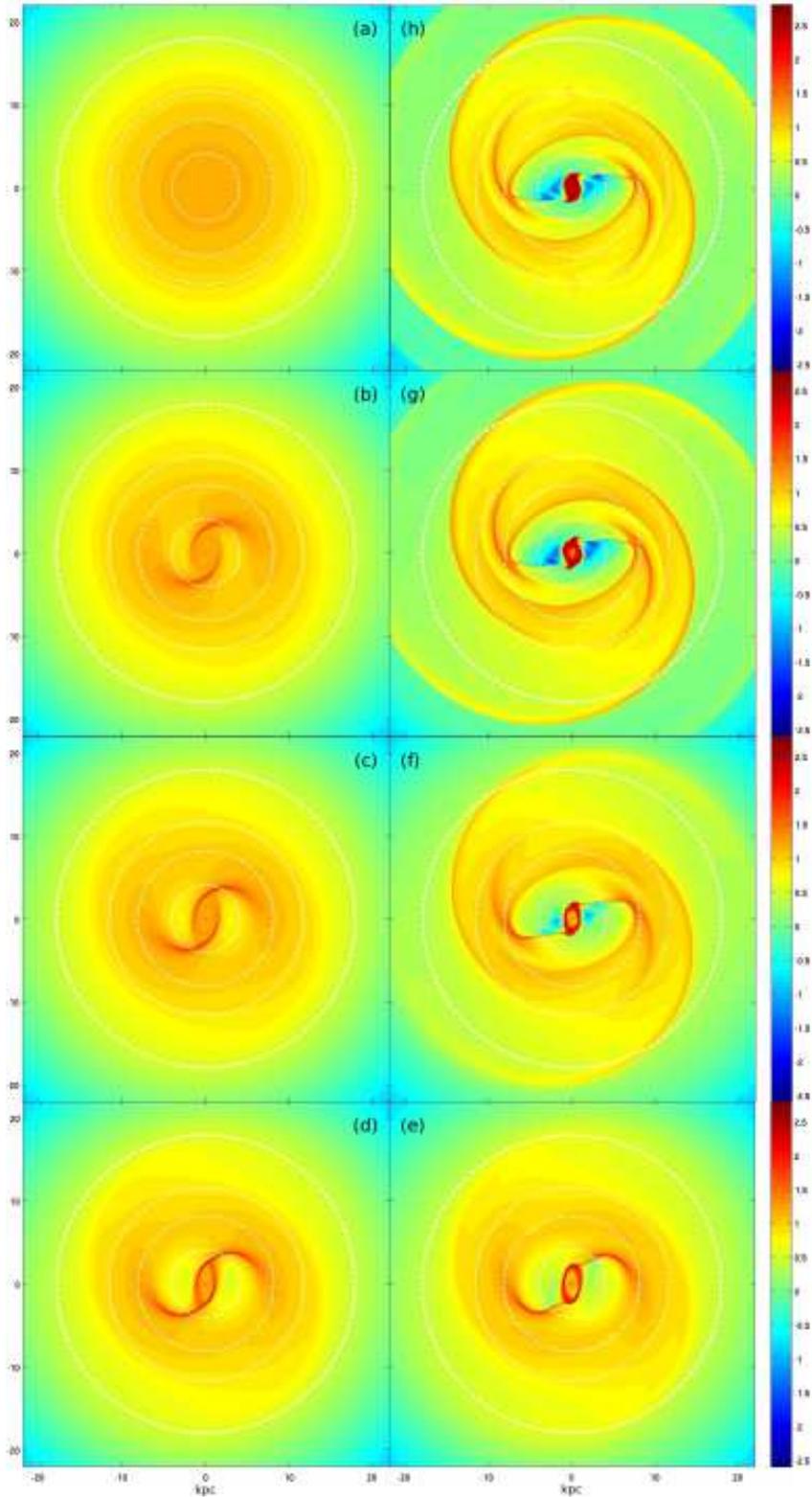}
\caption{Evolution of the bar-driven gaseous disk. The eight panels are arranged in counterclockwise order with increasing time.
The individual times are 0, 233, 273, 320, 371, 502, 711, 1237 Myr.
The bar is in horizontal orientation in each frame.
The color map denotes the surface density distribution in logarithmic scale.
There are four dashed circles indicating the positions of the OILR, the 4:1 UHR, the CR, and the OLR from the inside out in each frame.}
\label{evolution1097}
\end{figure}

\begin{figure}
\figurenum{8}
\epsscale{1.0}
\plotone{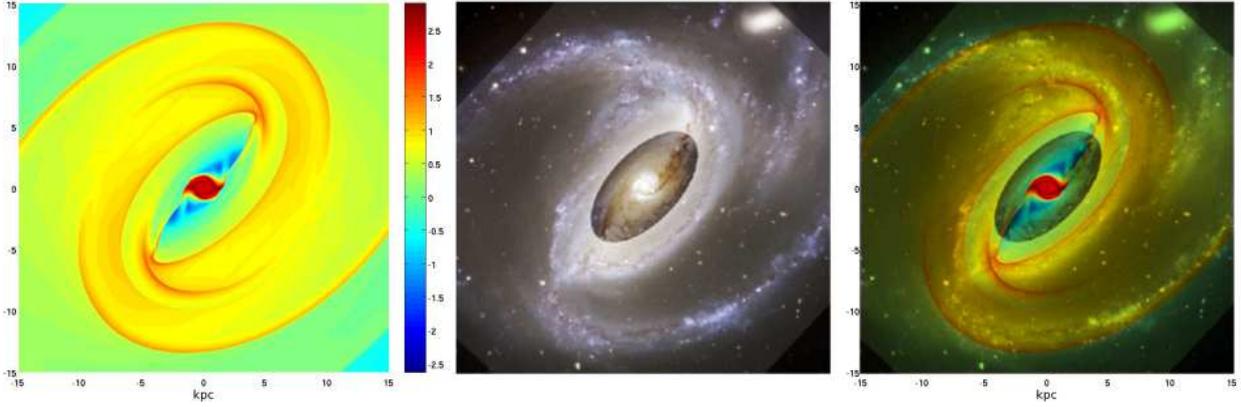}
\caption{Left panel: The projected surface density distribution of the simulation result in logarithmic scale 
in units of M$_\sun$ pc$^{-2}$.  The frame is taken after 4.35 revolutions of the bar, or 1237 Myr.
Middle panel: Replica of Figure~\ref{Oimage1097} with the brightness enhanced except for the central part.
Right panel: Superposition of the left two figures. North is at the top and east is to the left.
The bright nuclear starburst ring, the northwest off-centered dust lane, two lateral spiral arms along the bar region, as well as the southwest main spiral arm are reproduced by the simulation in both their shapes and locations.
All panels in this figure are on the same scale.}
\label{surfden1097}
\end{figure}

\begin{figure}
\figurenum{9}
\epsscale{1.0}
\plotone{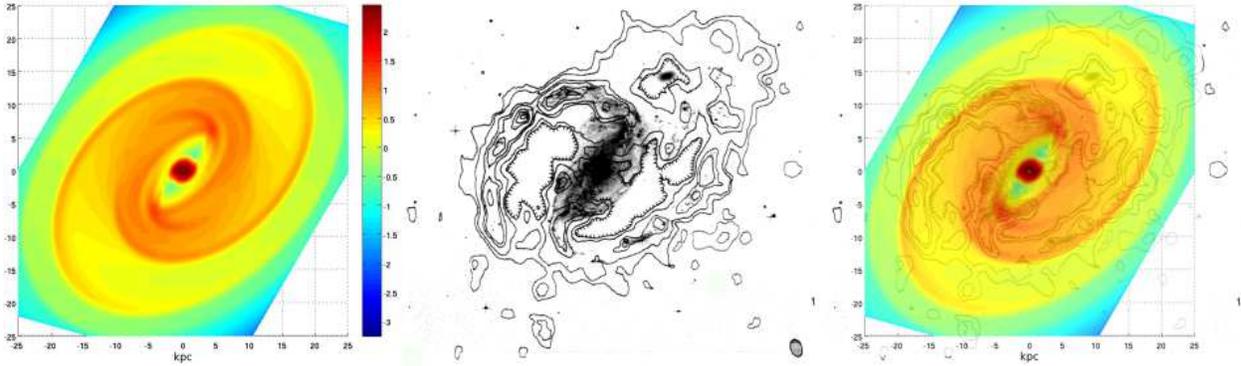}
\caption{Left panel: The projected simulated density distribution convolved with the synthesized beam shown in the middle panel.
The color map denotes the surface density distribution in logarithmic scale in units of M$_\sun$ pc$^{-2}$.
Middle panel: Contour plot of the HI column densities in NGC 1097 (Ondrechen et al. 1989) superposed on a photograph of the galaxy.
Contour levels are 4, 6, 10, 12, 15, and 17 $\times$ 10$^{20}$ cm $^{-2}$. 
Right panel: Superposition of the left two figures. North is at the top and east is to the left.
All panels in this figure are of the same scale.}
\label{surfden1097HI}
\end{figure}

\begin{figure}
\figurenum{10}
\epsscale{1.0}
\plotone{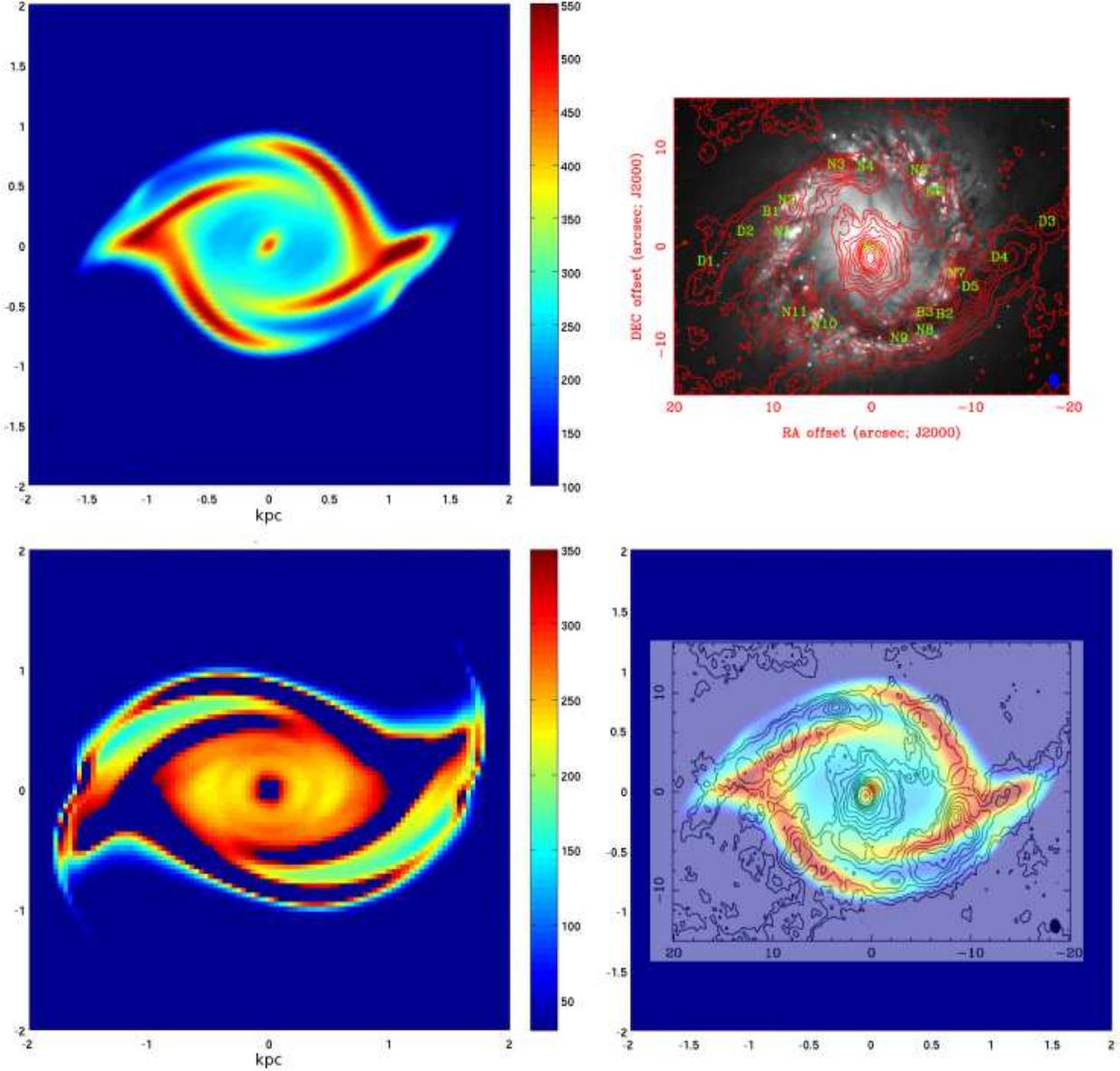}
\caption{Upper left: The central part of the projected simulated density map convolved with the synthesized beam 
in the upper right panel.  The color bar denotes the surface density on a linear scale in units of M$_\sun$ 
pc$^{-2}$.  Upper right: Replica of Figure~\ref{PYCOMT0}.  Lower right: The superposition of the upper left panel 
and the $^{12}$CO(J=2-1) emission (contours) in the upper right panel.  North is at the top and east is to the 
left.  The starburst ring in the simulation is of the same size as that in the observation.  The two groups of 
strongest emission knots in the northeast and the southwest parts of the starburst ring in the observation 
coincide with the densest regions in the simulation.  Lower left: The unprojected and unconvolved surface density 
map. To accentuate the weak spirals inside the ring, the high density region 
in the ring has been removed.  All panels in this figure are of the same scale.}
\label{surfden1097ctr}
\end{figure}

The central part of the projected simulated density map convolved with the synthesized beam in Figure~\ref{PYCOMT0} 
is shown in the upper left panel of Figure~\ref{surfden1097ctr}.  It is superimposed onto Figure~\ref{PYCOMT0} in 
the lower right panel of the same figure.  We have reproduced the starburst ring located at 800 pc from the 
central nucleus, which is comparable to the observation.  
We note that in the southwestern part of the ring, the knots corresponding to
the strongest emissions in the observation coincide with the densest region in the simulation. However, the match between our bisymmetric model and observation is not as good in the northeastern part of the ring, which is not symmetrical to the southwestern part in the observation. Nevertheless, in the northwestern and southeastern parts of the ring, two groups of knots corresponding to secondary strongest emissions in the observation do coincide with secondary densest regions in the simulation.
The HCN(J=3-2) intensity map in Figure 1 of \citet{hsieh12} shows features similar to those in the $^{12}$CO(J=2-1) map in Figure~\ref{PYCOMT0}.
However, since the beam size in the HCN map is about three times larger than that in the CO map, we only compare our simulation result with the latter.

In \citet{hsieh11}, the molecular clumps in their $^{12}$CO(J=2-1) observation are divided into three groups 
based on their locations and velocity dispersions.  The clumps located in the dust lanes are labeled D1,...,D5,
while those in the starburst ring are further distinguished by their velocity dispersion being 
greater or less than 30 km s$^{-1}$, and are named, respectively, B1,...,B3, and N1,...,N11 (see the upper right 
panel of Figure~\ref{surfden1097ctr}).  In the southwestern part of the ring, although N7 and D5 are close to 
each other, we can see in our simulation that D5 is on the convex side of the spiral while N7 is on the concave side of the 
spiral and is in the ring.  
This comparison supports the interpretation that N7 and D5 belong to different 
components and are characterized by different properties in the observation.  In Section 3.4.1 of \citet{hsieh11}, the density 
of narrow line clumps is found to be (5.3 $\pm$ 4.2) times higher than that of broad line clumps.  Measuring the 
positions of B2 and N9 in our simulation, we find B2 is located in the lower density region 
between two spirals, while N9 is located in the higher density region on the ring.  The surface densities 
in these two regions differ by a factor of about 2. 
Since the signal 
to noise ratio is lower in the north-east part in the $^{12}$CO(J=2-1) observation, it is difficult to make similar
comparisons in this region.
 
Although the gaseous distribution in the nuclear ring in our simulation is not as clumpy as that in the observation,
the comparison, here, mainly shows that in models including self-gravity of the gas, gravitational instabilities can result in a richer and more realistic set of features than in simulations without self-gravity.  
Previous studies show that 
clumpy features can be generated by including additional mechanisms (e.g.,
heating and cooling or star formation) in simulations \citep[e.g.,][]{lindblad96b, wada08, agertz11},
however, we concentrate on the effect of self-gravity in this study.

For the observed circumnuclear disk, four emission sites extend outward from the disk.  The southeast protruding 
emission is connected with the starburst ring and, if we neglect the northeast feature, the other three are in 
alignment parallel to the direction of the dust lanes.  In the upper left panel of Figure~\ref{surfden1097ctr}, the 
central high density disk also connects with the ring in the direction parallel to that of dust lanes.  However, 
in the unprojected density map, which is not convolved with the synthesized beam, it is seen that the central 
disk and the ring are partially connected by a pair of tightly wound spirals.  These features are separated from 
the central disk and the ring and only appear to be connected after convolution.
 Through this pair of tightly wound spirals which penetrate the nuclear starburst ring, a portion of the gas can flow into the central region inside the ring to form and maintain the circumnuclear disk.
However, since the resolution of our simulation is 39 $\times$ 39 pc$^{2}$, we do not compare our simulation results with observations on the 200 pc scale central region of the galaxy in, for example, \citet{hicks09} and \citet{davies09}.

\subsubsection{Velocity fields}
\begin{figure}
\figurenum{11}
\epsscale{1.0}
\plotone{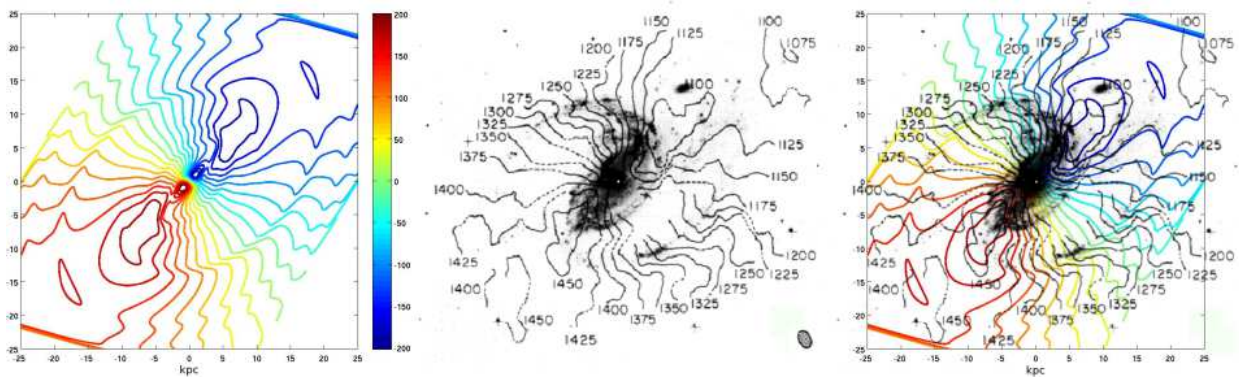}
\caption{Comparisons of simulated and observed velocity fields.
Left panel: Isovelocity curves from the simulated velocity field convolved with the synthesized beams of the observations of Ondrechen et al. (1989) .
Middle panel: Replica of Figure~\ref{HIvf}.
Right panel: The superposition of the left two.
North is at the top and east is to the left.
The velocity contours bend inwards regularly along the southwest spiral arms in both the simulation and observation.
All panels in this figure are of the same scale.} 
\label{HIvfsimobs1097}
\end{figure}

The isovelocity contours of the simulated velocity field, as obtained after convolution with the synthesized beam 
of the observation, are plotted on top of the observed HI velocity field in the left panel of 
Figure~\ref{HIvfsimobs1097}.  We note that the velocity contours in the southwest arm bend inwards regularly 
along the spiral arms in both the simulation and observation as illustrated in the right panel of 
Figure~\ref{HIvfsimobs1097}.  This is one of the primary distinctive features of the density waves excited at the OLR \citep{yuankuo98}.
  However, this feature is not obviously present in the northeast spiral arm in the observation, which may be 
due to an influence by the gravitational field of NGC 1097A on the northeast spiral arm.

\begin{figure}
\figurenum{12}
\epsscale{1.0}
\plotone{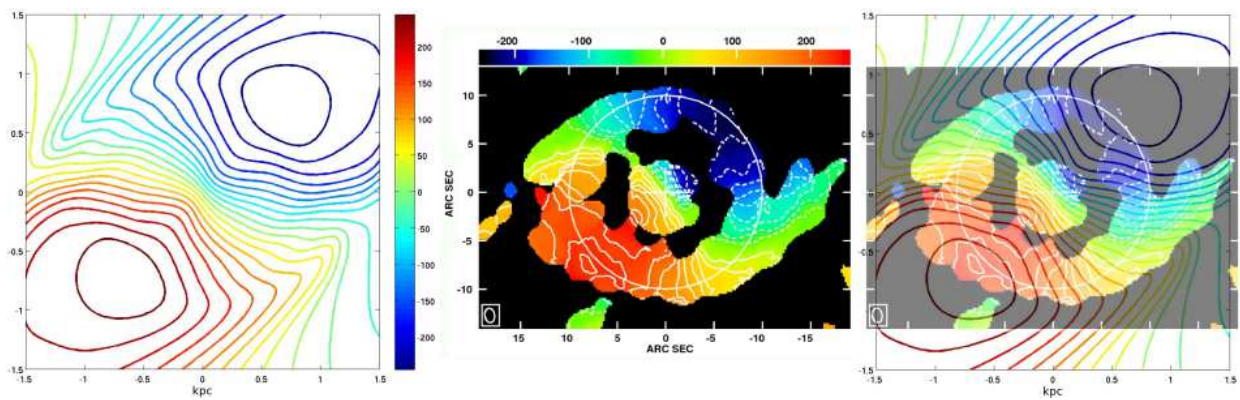}
\caption{Left panel: The central part of the isovelocity curves from the simulated velocity field convolved with the synthesized beam of the $^{12}$CO (J = 2-1) observation of \citet{hsieh11}.
Middle panel: Replica of Figure~\ref{COvf}.
Right panel: Superposition of the left two.
The overall velocity gradient of the central molecular disk and the starburst ring is reproduced in our simulation.
All panels in this figure are of the same scale.}
\label{COvfsimobs1097}
\end{figure}

A comparision of the central part of the simulated velocity fields with that of the $^{12}$CO (J = 2-1) observation 
is shown in Figure~\ref{COvfsimobs1097}. It can be seen that the central part of the galaxy is less influenced by 
the companion galaxy and that the simulation result from the bisymmetrical model matches well with the observation.
In particular, the overall velocity gradient of the central molecular disk and the starburst ring is reproduced 
in our simulation.  The numerical results reveal that gas motion in the region inside of the molecular ring is dominated by circular 
motion.
However, in the low resolution observation of $^{12}$CO (J=2-1) in \citet{hsieh08}, the isovelocity 
contours of the velocity map exhibit an S-shape structure with the end of the S-shape nearly parallel to the 
dust lanes.  This structure, which is even more prominent in the $^{12}$CO (J=1-0) velocity map of \citet{kohno03}, indicates the presence of noncircular motion \citep[e.g.,][]{kalnajs78} in the regions where the dust lanes 
connect with the molecular ring.  Such an S-shape structure is also reproduced by our simulation.

The red line in Figure~\ref{rotc1097} is the evolved rotation curve along the major axis in our simulation, which shows a large deviation from the initial rotation curve within 4 kpc produced by the bar perturbation.
It can qualitatively reproduce both the fast-rotating feature of the CO rotation curve near 1 kpc and the drop shown by the HI data outside the starburst ring, which are also seen in \citet{dicaire08}.
Between 4 kpc and 8 kpc, the rotation curves from different observations show a large diversity.
Since the gas density in this region, which roughly corresponds to the area outside the starburst ring and inside the bar region, is low except for the area along the dust lanes and there are strong shocks across the dust lanes, the uncertainties of the derived rotation curve from observations could be quite large.
Therefore, one should refer to 2-D velocity fields when comparing observations and simulation results in this region.
Although the evolved simulated rotation curve outside the 15 kpc is lower than that of the HI observation, it matches well with the H$\alpha$ observation by \citet{dicaire08}.

\subsection{Gravitational instability and the circumnuclear ring}
It has been known for some time that there exists a threshold in the gas surface density within each galaxy below which little star formation could occur.  
On kpc scales, the \ion{H}{1} density seems to serve as a good indicator for this density threshold \citep{davies76,hunter86,guiderdoni87,skillman87,hulst87}.  
Recent higher-resolution observations, though, has posed some questions on this notion.  \citet{bigiel08} reported that the \ion{H}{1} density tends to saturate on smaller scales and this saturated density may manifest itself as a threshold for star formation.
 They further reported that the FUV data may not change as steeply around a \ion{H}{1} density threshold as previously found in H$\alpha$ data \citep{bigiel10}.
The origin of this apparent contradiction is not clear and more analyses are required to resolve this issue.

In spite of this situation, the phenomenon of density threshold for star formation is often associated with large-scale gravitational instabilities \citep{kennicutt89}. 
 However, the simple Toomre stability parameter for the gas, $Q \equiv a \kappa / \pi G \sigma$, where $a$, $\kappa$, $G$
and $\sigma$ are the sound speed of the gas, the epicycle frequency, the gravitational constant,
and the gas surface density, respectively, is not robust in predicting star formation \citep[e.g.,][]{martin01,wong02}.  Furthermore, the stellar component of a galactic disk 
\citep[e.g.,][]{yang07,leroy08} and the atomic-to-molecular transition 
\citep[e.g.,][]{schaye04}
may be also important in contributing to the gravitational instability of the disk.

\begin{figure}
\figurenum{13}
\epsscale{1.0}
\plotone{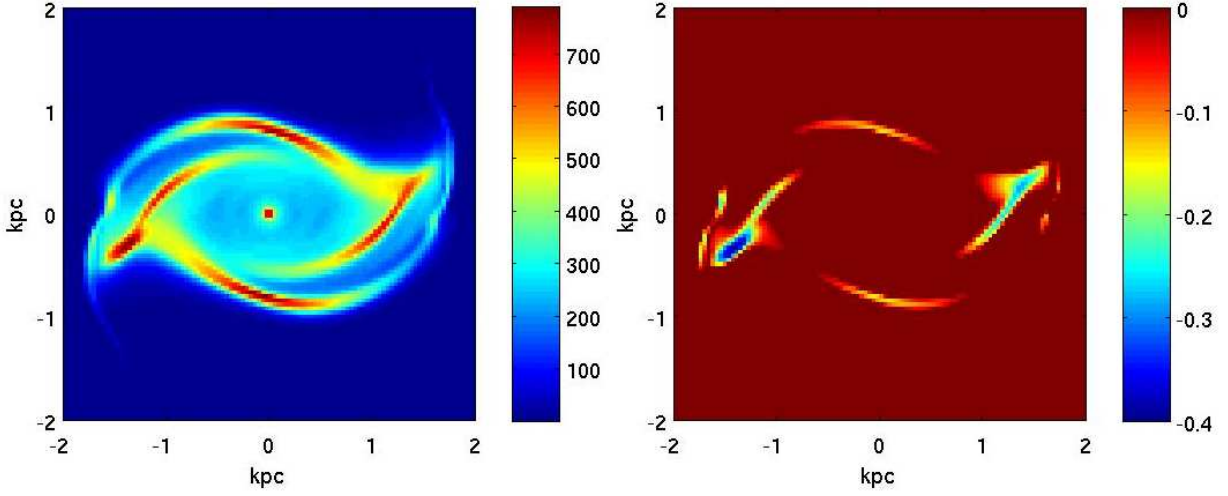}
\caption{Left panel: The simulated face-on density map of the central part of the galaxy.
Right panel: The Toomre Q map in logarithmic scale. 
The Q-values of the two densest regions on the ring are both lower than 1.}
\label{Qvalue}
\end{figure}

Although our simple model for NGC~1097 cannot address the complicated dynamics associated with star-gas interactions and atomic-to-molecular transition, our results are consistent with observations,
which shows that the star-forming circumnuclear ring is located in the gravitationally unstable region \citep{hsieh11}.
In our simulations, gravitational instability appears at the locations where the dust lanes connect with the circumnuclear ring 
when the ring has just been formed ($\sim$1.5 bar revolutions).
During the subsequent evolution of the gaseous disk, the instability propagates within the ring,
leading to complex and asymmetrical structures.
In Figure~\ref{Qvalue}, a map of the Toomre $Q$-parameter is shown for the central 2 kpc region of the galaxy at the time selected for comparison of the simulation result with observations.
The left panel corresponds to the map of the surface density and the right panel illustrates the $Q$ map on a logarithmic scale.
The $Q$-parameter of the two densest regions (blue color in the right panel) on the ring are both less than one.
These regions coincide with the N1, N2, N7, N9, and B1--B3 molecular clumps reported in \citet{hsieh11}.
The second densest regions on the ring situated on the major axis of the bar are also gravitationally unstable and coincide with the N5, N6, N10, and N11 molecular clumps identified in \citet{hsieh11}.
Therefore, it seems that in the case of NGC~1097, gravitational instability of the gas disk itself is sufficient in predicting star forming sites in the circumnuclear region.

\subsection{Mass inflow rate}

\begin{figure}
\figurenum{14}
\epsscale{1.0}
\plotone{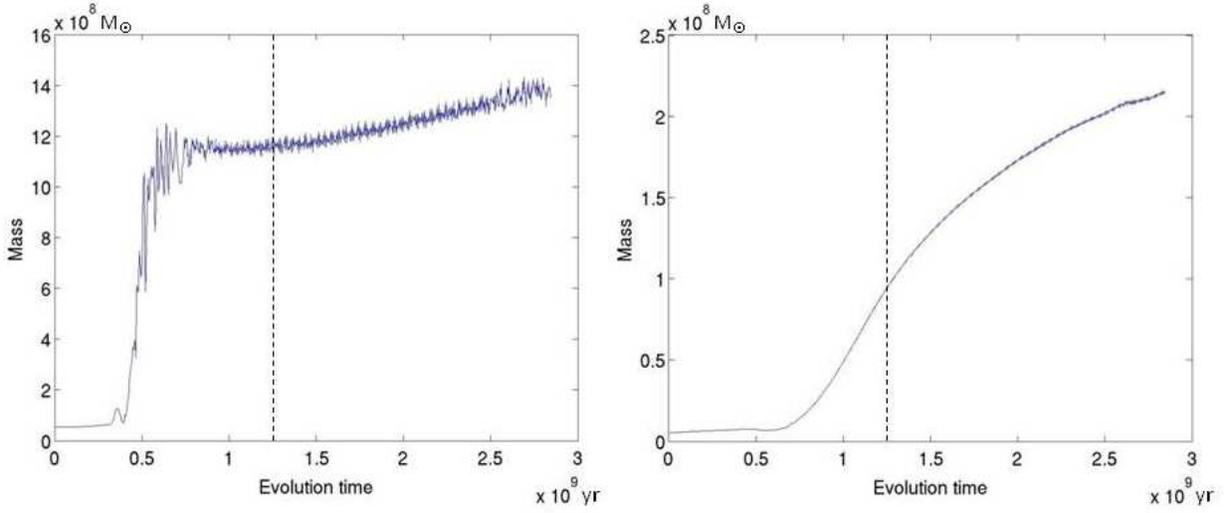}
\caption{Left panel: The time evolution of the total gaseous mass in the starburst ring.
The black dashed line indicates the time we select for comparision of the simulation result with observations.
The mass at that time is 11.6 $\times$  10$^{8}$  M$_\sun$.
Right panel: The total gaseous mass inside the starburst ring as a function of time.
The mass inflow rate at the time we select is 0.17 M$_\sun$ yr $^{-1}$. 
The values of the mass of the circumnuclear disk estimated by \citet{hsieh08} and obtained by our simulation are 6.5 $\times$ 10$^{7}$ and 9.47 $\times$ 10$^{7}$ M$_\sun$ respectively.} 
\label{massinflow}
\end{figure}

The left and right panels of Figure~\ref{massinflow} show the time evolution of the total gaseous mass in the 
simulated starburst ring and interior to this ring respectively. The delay in the formation of the ring (after 
500 Myr) is attributed to the start up procedure in which the strength of the bar potential is gradually 
increased during the first two revolutions of the bar. 
Once formed, its total mass increases with time until 
750 Myr, whereupon it remained nearly constant at a level of $11.3 \times 10^8 
M_{\odot}$ until 1.1 Gyr. Thereafter, the mass in the ring increases, but at a slower rate than during the initial
build-up.  In contrast, the mass interior to the ring increases continuously, with the time rate of change greater 
during the initial phase from 750 Myr to 1.3 Gyr than during the time after 1.3 Gyr. 

At a time corresponding to the dashed line in the left panel of Figure~\ref{massinflow}, which we select for 
comparison with observations, the mass in the starburst ring in the simulation is 11.6 $\times$ 10$^{8}$ 
M$_\sun$.  This is to be compared with the estimated value from observation of 5.8 $\pm$ 0.6 $\times$ 10$^{8}$ 
M$_\sun$ \citep{hsieh08}.  The estimated surface density star formation rates of the molecular clumps 
situated on the starburst ring range from 0.58 $\pm$ 0.08 to 4.07 $\pm$ 0.34 M$_\sun$ yr$^{-1}$ kpc$^{-2}$ 
\citep{hsieh11}.  This yields an estimated average star formation rate on the starburst ring of 3.1 M$_\sun$ yr$^{-1}$. 

\citet{elmegreen96} investigated star formation scaling laws and found that the duration of star formation
in a region of size L tends to increase as L$^{1/2}$.  For a region characterized by $\sim$ 1 kpc, the time 
scale for star formation is $\sim$ 30 Myr.  
\citet{elmegreen97} suggested the starburst process in nuclear rings in galaxies 
consists of three phases. In particular, the gas density first increases above the local virial density, which 
is followed by star formation taking place at high density.  In the final phase, the gas density decreases below the virial 
density.  Since the virial density in nuclear regions in galaxies is always very high, star formation is always intense and 
burst like.  The consumption time is short and once the gas 
density decreases below the virial density,  star formation ends quickly with the next phase delayed until 
the gas density is increased above the critical value again.  For NGC 1097, the size of the starburst ring is 
$\sim$ 1.4 kpc and the estimated duration for a high star formation rate (3.1 M$_\sun$ yr$^{-1}$) is $\sim$ 35 Myr.

The mass inflow rate was calculated from the slope of the total mass curve during the period between 
1.1 Gyr and 1.6 Gyr, resulting in a rate of 0.11 M$_\sun$ yr$^{-1}$.  Although this is much lower than the 
observed star formation rate, the time scale of the mass inflow in our model is much longer than the duration of the starburst 
on the ring. 
\citet{sandstrom10} has estimated that the lifespans of the star clusters on the starburst ring are between 1$\sim$ 10 Myr.
\citet{sakamoto99} have also mentioned that the current mass inflow rate is not required to be larger than the current star formation rate.
As long as the total amount of gas transported to the central kiloparsec of barred galaxies is larger than the total amount of gas consumed by star formation, 
we should see high gas concentration in the central regions of the barred galaxies.

In the right panel of Figure~\ref{massinflow}, it can be seen that the mass interior to the ring at our 
selected epoch (the dashed line) is 9.47 $\times$ 10$^{7}$ M$_\sun$, which is comparable to the mass of 
the circumnuclear disk (6.5 $\times$ 10$^{7}$ M$_\sun$) estimated from observation \citep{hsieh08} .
The rate of mass accumulation in this region can be estimated from the slopes of the total mass curve 
between 1 Gyr and 1.3 Gyr.  
It is found to be 0.17 M$_\sun$ yr$^{-1}$.  

In \citet{nemmen12}, the dimensionless mass accretion rate is defined as $\dot{m} = \dot{M}/\dot{M}_{Edd}$, where $M$ is the black hole mass
and the Eddington accretion rate ($\dot{M}$) is equal to $22M/(10^{9}M_\sun) M_\sun$yr$^{-1}$.
For NGC 1097, the black hole mass is $(1.2 \pm 0.2) \times 10^{8} M_\sun$ \citep{lewis06} whereas
$\dot{m}$ is $6.4 \times 10^{-3}$ \citep{nemmen12}.
Therefore, $\dot{M}$ equals to 0.017 M$_\sun$ yr$^{-1}$.
This accretion rate is inferred
at the outer radius of the Advection-dominated accretion flow, which for NGC 1097 in \citet{nemmen12} is 225 times the Schwarzschild radius or
about 530 AUs.
Although the mass inflow rate in our model is 
higher than the observed mass accretion rate for the AGN, the scales are very different.
From a few hundred parsecs to the AU scale nucleus, the mass inflow rate could be reduced 
as other mechanisms may operate in transporting gas into the nucleus from the 
circumnuclear disk \citep{shlosman89}. 
However, we note that 
the mass inflow in our model lasts for more than 2 Gyr and can sustain the circumnuclear disk to supply enough mass into the AU scale nucleus.

Since the expected star formation efficiencies for entire giant molecular clouds (GMCs)  are typically only 1-5\% \citep{lada03}, and the structures we are dealing with here are larger than GMCs, we expect the effect of star formation or stellar mass loss processes on our mass inflow rate estimates should not be significant.

\section{Concluding remarks}
We have performed two-dimensional modeling of the barred spiral galaxy NGC 1097 using the Antares hydrodynamics code coupled with the Poisson equation to include the self-gravity of the gas disk in the calculation.
For a perturbing bar potential (Eqs. 2 and 3) described by the angular speed $\Omega_p$, the location of the potential minimum a$_{1}$, and the strength of the bar $f_{OILR}$, we show that the primary density and velocity features of the galaxy can be reproduced provided that the parameters are in the range of  $\Omega_p$ = 21.6 km s$^{-1}$ kpc$^{-1}$, a$_{1}$ = 2.2 kpc, and $f_{OILR}$ = 21\%.
Specifically, with these parameters, the observed shapes and positions of the circumnuclear starburst ring, the northwest dust lane, the two lateral spiral arms along the bar region as well as the southwest main spiral arm are well reproduced in our model.
In addition, our simulation supports the view 
that the resolved molecular clumps at the giant molecular cloud association (GMA) scale of 200-300 pc in the starburst ring by \citet{hsieh11} are characterized by different properties.
Furthermore, the connection between the circumnuclear disk and the starburst ring in the observation 
may be due to the
unresolved tightly wound spirals which are separated from the central disk 
and the ring in the simulation before convolution with the synthesized beam of the observation.
Through this pair of tightly wound spirals which penetrate the nuclear starburst ring, a portion of the gas can flow into the central region inside the ring to form and maintain the circumnuclear disk.
  
On comparison with the observed velocity fields, the velocity contours in the southwest main spiral arm bend inwards regularly along the spiral arms in both the simulation and observation.
Although this feature is not obvious in the northeast main spiral arm, which may be 
influenced by the companion galaxy NGC 1097A,
it is still indicative that the main spiral arms 
can be attributed to the density waves excited at the OLR by the bar potential.

By including the self-gravity of the gas disk in our calculation, 
we have studied the gravitational stability and the mass inflow in the gas disk.
We find that the starburst ring is gravitationally 
unstable in our simulation, which is consistent with the result in \citet{hsieh11}.
This comparison indicates that the Toomre $Q$ parameter is a useful diagnostic of star forming sites in the case of NGC 1097.

The mass inflow rate for the starburst ring is 0.11 M$_\sun$ yr$^{-1}$, lasting
for more than 1.5 Gyr in our simulation.
Although the current mass inflow rate in the 
simulation is lower than the observed current star formation rate (3.1 M$_\sun$ yr$^{-1}$),
the time scale of the mass inflow in our model is much longer than the duration of the starburst on the ring.

The accretion rate for the supermassive black hole of NGC 1097 estimated from observation is 0.017 M$_\sun$ yr$^{-1}$. 
This is lower than the inflow rate of the gas entering the central region within the starburst ring in our simulation (0.17 M$_\sun$ yr$^{-1}$). 
However, the mass inflow rate in our model is evaluated at 
the scale of a few hundred parsecs while the spatial scale of the 
observed mass accretion rate for the AGN is only a few hundred AUs.  
Our results do not 
address the mass flow in the inner 10 pc, but it is likely that AGN activity as well as 
stellar and gas dynamical mechanisms are important in influencing the gas flow in the 
nucleus. 
Nevertheless, the mass inflow in our model lasts for more than 2 Gyr, which should 
be able to sustain the circumnuclear disk in supplying sufficient  mass into the AU-scale nucleus.

\acknowledgments
We thank the referee for his/her comments which helped to significantly improve  the clarity and presentation of this work.
We also thank Mr. Sam Tseng for assistance on the computational facilities and resources.

\clearpage

\begin{deluxetable}{ll}
\tabletypesize{\scriptsize}
\tablewidth{0pc}
\tablecaption{Properties of NGC 1097}
\tablehead{
\colhead{Parameter} &
\colhead{Value} 
}
\startdata
$\alpha(J2000)$\tablenotemark{a} & 02$^h$ 46$^m$ 16\rlap{.}$^s$96\\
$\delta(J2000)$\tablenotemark{a} & $-$30$^{\circ}$ 16$^{\prime}$ 28$^{\prime\prime}$.9\\
Morphology\tablenotemark{b} & SB(s)b \\
Nuclear activity\tablenotemark{c} & Type 1 Seyfert \\
Line of nodes\tablenotemark{d} & 134$^{\circ} \pm$ 3$^{\circ}$ \\
Inclination\tablenotemark{d} & 46$^{\circ} \pm$ 5$^{\circ}$ \\
P.A. of bar\tablenotemark{e} & 148$^{\circ} \pm$ 3$^{\circ}$ \\
Adopted distance & 16.96 Mpc \\
\enddata
\tablenotetext{a}{Hummel et al. (1987)}
\tablenotetext{b}{de Vaucouleurs et al. (1991)}
\tablenotetext{c}{Storchi-Bergmann et all (1993)}
\tablenotetext{d}{Ondrechen et al. (1989)}
\tablenotetext{e}{Quillen et al. (1995)}
\end{deluxetable}

\clearpage

\clearpage
\bibliography{ngc1097ref}

\begin{thebibliography}{71}
\expandafter\ifx\csname natexlab\endcsname\relax\def\natexlab#1{#1}\fi

\bibitem[{{Agertz} {et~al.}(2011){Agertz}, {Teyssier}, \& {Moore}}]{agertz11}
{Agertz}, O., {Teyssier}, R., \& {Moore}, B. 2011, \mnras, 410, 1391

\bibitem[{{Athanassoula}(1992)}]{athan92}
{Athanassoula}, E. 1992, \mnras, 259, 345

\bibitem[{{Barth} {et~al.}(1995){Barth}, {Ho}, {Filippenko}, \&
  {Sargent}}]{barth95}
{Barth}, A.~J., {Ho}, L.~C., {Filippenko}, A.~V., \& {Sargent}, W.~L. 1995,
  \aj, 110, 1009

\bibitem[{{Bigiel} {et~al.}(2010){Bigiel}, {Leroy}, {Walter}, {Blitz},
  {Brinks}, {de Blok}, \& {Madore}}]{bigiel10}
{Bigiel}, F., {Leroy}, A., {Walter}, F., {et~al.} 2010, \aj, 140, 1194

\bibitem[{{Bigiel} {et~al.}(2008){Bigiel}, {Leroy}, {Walter}, {Brinks}, {de
  Blok}, {Madore}, \& {Thornley}}]{bigiel08}
---. 2008, \aj, 136, 2846

\bibitem[{{Boone} {et~al.}(2004){Boone}, {Garc{\'{\i}}a-Burillo}, {Combes},
  {Hunt}, {Baker}, {Tacconi}, {Eckart}, {Neri}, {Leon}, {Schinnerer}, \&
  {Englmaier}}]{boone04}
{Boone}, F., {Garc{\'{\i}}a-Burillo}, S., {Combes}, F., {et~al.} 2004, in
  Astronomical Society of the Pacific Conference Series, Vol. 320, The Neutral
  ISM in Starburst Galaxies, ed. S.~{Aalto}, S.~{Huttemeister}, \& A.~{Pedlar},
  269

\bibitem[{{Crosthwaite}(2002)}]{crosthwaite02}
{Crosthwaite}, L.~P. 2002, \pasp, 114, 929

\bibitem[{{Davies} {et~al.}(1976){Davies}, {Elliott}, \& {Meaburn}}]{davies76}
{Davies}, R.~D., {Elliott}, K.~H., \& {Meaburn}, J. 1976, \memras, 81, 89

\bibitem[{{Davies} {et~al.}(2009){Davies}, {Maciejewski}, {Hicks}, {Tacconi},
  {Genzel}, \& {Engel}}]{davies09}
{Davies}, R.~I., {Maciejewski}, W., {Hicks}, E.~K.~S., {et~al.} 2009, \apj,
  702, 114

\bibitem[{{de Vaucouleurs} {et~al.}(1991){de Vaucouleurs}, {de Vaucouleurs},
  {Corwin}, {Buta}, {Paturel}, \& {Fouque}}]{vaucouleurs91}
{de Vaucouleurs}, G., {de Vaucouleurs}, A., {Corwin}, Jr., H.~G., {et~al.}
  1991, \skytel, 82, 621

\bibitem[{{Dicaire} {et~al.}(2008){Dicaire}, {Carignan}, {Amram}, {Hernandez},
  {Chemin}, {Daigle}, {de Denus-Baillargeon}, {Balkowski}, {Boselli}, {Fathi},
  \& {Kennicutt}}]{dicaire08}
{Dicaire}, I., {Carignan}, C., {Amram}, P., {et~al.} 2008, \mnras, 385, 553

\bibitem[{{Elmegreen}(1997)}]{elmegreen97}
{Elmegreen}, B.~G. 1997, in Revista Mexicana de Astronomia y Astrofisica, vol.
  27, Vol.~6, Revista Mexicana de Astronomia y Astrofisica Conference Series,
  ed. J.~{Franco}, R.~{Terlevich}, \& A.~{Serrano}, 165

\bibitem[{{Elmegreen} \& {Efremov}(1996)}]{elmegreen96}
{Elmegreen}, B.~G., \& {Efremov}, Y.~N. 1996, \apj, 466, 802

\bibitem[{{Elmegreen} \& {Elmegreen}(1990)}]{elmegreen90}
{Elmegreen}, B.~G., \& {Elmegreen}, D.~M. 1990, \apj, 355, 52

\bibitem[{{Garc{\'{\i}}a-Burillo} {et~al.}(2005){Garc{\'{\i}}a-Burillo},
  {Combes}, {Schinnerer}, {Boone}, \& {Hunt}}]{garcia05}
{Garc{\'{\i}}a-Burillo}, S., {Combes}, F., {Schinnerer}, E., {Boone}, F., \&
  {Hunt}, L.~K. 2005, \aap, 441, 1011

\bibitem[{{Garc{\'{\i}}a-Burillo} {et~al.}(2004){Garc{\'{\i}}a-Burillo},
  {Combes}, {Schinnerer}, {Boone}, {Hunt}, {Eckart}, {Tacconi}, {Leon},
  {Baker}, {Englmaier}, \& {Neri}}]{garcia04}
{Garc{\'{\i}}a-Burillo}, S., {Combes}, F., {Schinnerer}, E., {et~al.} 2004, in
  IAU Symposium, Vol. 222, The Interplay Among Black Holes, Stars and ISM in
  Galactic Nuclei, ed. T.~{Storchi-Bergmann}, L.~C. {Ho}, \& H.~R. {Schmitt},
  427--430

\bibitem[{{Guiderdoni}(1987)}]{guiderdoni87}
{Guiderdoni}, B. 1987, \aap, 172, 27

\bibitem[{{Hicks} {et~al.}(2009){Hicks}, {Davies}, {Malkan}, {Genzel},
  {Tacconi}, {M{\"u}ller S{\'a}nchez}, \& {Sternberg}}]{hicks09}
{Hicks}, E.~K.~S., {Davies}, R.~I., {Malkan}, M.~A., {et~al.} 2009, \apj, 696,
  448

\bibitem[{{Higdon} \& {Wallin}(2003)}]{higdon03}
{Higdon}, J.~L., \& {Wallin}, J.~F. 2003, \apj, 585, 281

\bibitem[{{Ho} {et~al.}(2004){Ho}, {Moran}, \& {Lo}}]{ho04}
{Ho}, P.~T.~P., {Moran}, J.~M., \& {Lo}, K.~Y. 2004, \apjl, 616, L1

\bibitem[{{Hsieh} {et~al.}(2012){Hsieh}, {Ho}, {Kohno}, {Hwang}, \&
  {Matsushita}}]{hsieh12}
{Hsieh}, P.-Y., {Ho}, P.~T.~P., {Kohno}, K., {Hwang}, C.-Y., \& {Matsushita},
  S. 2012, \apj, 747, 90

\bibitem[{{Hsieh} {et~al.}(2008){Hsieh}, {Matsushita}, {Lim}, {Kohno}, \&
  {Sawada-Satoh}}]{hsieh08}
{Hsieh}, P.-Y., {Matsushita}, S., {Lim}, J., {Kohno}, K., \& {Sawada-Satoh}, S.
  2008, \apj, 683, 70

\bibitem[{{Hsieh} {et~al.}(2011){Hsieh}, {Matsushita}, {Liu}, {Ho}, {Oi}, \&
  {Wu}}]{hsieh11}
{Hsieh}, P.-Y., {Matsushita}, S., {Liu}, G., {et~al.} 2011, \apj, 736, 129

\bibitem[{{Hummel} {et~al.}(1987){Hummel}, {van der Hulst}, \&
  {Keel}}]{hummel87}
{Hummel}, E., {van der Hulst}, J.~M., \& {Keel}, W.~C. 1987, \aap, 172, 32

\bibitem[{{Hunter} \& {Gallagher}(1986)}]{hunter86}
{Hunter}, D.~A., \& {Gallagher}, III, J.~S. 1986, \pasp, 98, 5

\bibitem[{{Kalnajs}(1978)}]{kalnajs78}
{Kalnajs}, A.~J. 1978, in IAU Symposium, Vol.~77, Structure and Properties of
  Nearby Galaxies, ed. E.~M. {Berkhuijsen} \& R.~{Wielebinski}, 113--125

\bibitem[{{Keel}(1983)}]{keel83}
{Keel}, W.~C. 1983, \apj, 269, 466

\bibitem[{{Kenney} {et~al.}(1992){Kenney}, {Wilson}, {Scoville}, {Devereux}, \&
  {Young}}]{kenney92}
{Kenney}, J.~D.~P., {Wilson}, C.~D., {Scoville}, N.~Z., {Devereux}, N.~A., \&
  {Young}, J.~S. 1992, \apjl, 395, L79

\bibitem[{{Kennicutt}(1989)}]{kennicutt89}
{Kennicutt}, Jr., R.~C. 1989, \apj, 344, 685

\bibitem[{{Knapen} {et~al.}(1995){Knapen}, {Beckman}, {Heller}, {Shlosman}, \&
  {de Jong}}]{knapen95}
{Knapen}, J.~H., {Beckman}, J.~E., {Heller}, C.~H., {Shlosman}, I., \& {de
  Jong}, R.~S. 1995, \apj, 454, 623

\bibitem[{{Kohno} {et~al.}(2003){Kohno}, {Ishizuki}, {Matsushita},
  {Vila-Vilar{\'o}}, \& {Kawabe}}]{kohno03}
{Kohno}, K., {Ishizuki}, S., {Matsushita}, S., {Vila-Vilar{\'o}}, B., \&
  {Kawabe}, R. 2003, \pasj, 55, L1

\bibitem[{{Koribalski} {et~al.}(2004){Koribalski}, {Staveley-Smith}, {Kilborn},
  {Ryder}, {Kraan-Korteweg}, {Ryan-Weber}, {Ekers}, {Jerjen}, {Henning},
  {Putman}, {Zwaan}, {de Blok}, {Calabretta}, {Disney}, {Minchin}, {Bhathal},
  {Boyce}, {Drinkwater}, {Freeman}, {Gibson}, {Green}, {Haynes}, {Juraszek},
  {Kesteven}, {Knezek}, {Mader}, {Marquarding}, {Meyer}, {Mould}, {Oosterloo},
  {O'Brien}, {Price}, {Sadler}, {Schr{\"o}der}, {Stewart}, {Stootman}, {Waugh},
  {Warren}, {Webster}, \& {Wright}}]{koribalski04}
{Koribalski}, B.~S., {Staveley-Smith}, L., {Kilborn}, V.~A., {et~al.} 2004,
  \aj, 128, 16

\bibitem[{{Kotilainen} {et~al.}(2000){Kotilainen}, {Reunanen}, {Laine}, \&
  {Ryder}}]{kotilainen00}
{Kotilainen}, J.~K., {Reunanen}, J., {Laine}, S., \& {Ryder}, S.~D. 2000, \aap,
  353, 834

\bibitem[{{Lada} \& {Lada}(2003)}]{lada03}
{Lada}, C.~J., \& {Lada}, E.~A. 2003, \araa, 41, 57

\bibitem[{{Leroy} {et~al.}(2008){Leroy}, {Walter}, {Brinks}, {Bigiel}, {de
  Blok}, {Madore}, \& {Thornley}}]{leroy08}
{Leroy}, A.~K., {Walter}, F., {Brinks}, E., {et~al.} 2008, \aj, 136, 2782

\bibitem[{{Lewis} \& {Eracleous}(2006)}]{lewis06}
{Lewis}, K.~T., \& {Eracleous}, M. 2006, \apj, 642, 711

\bibitem[{{Lin} {et~al.}(2011){Lin}, {Taam}, {Yen}, {Muller}, \& {Lim}}]{lin11}
{Lin}, L.-H., {Taam}, R.~E., {Yen}, D.~C.~C., {Muller}, S., \& {Lim}, J. 2011,
  \apj, 731, 15

\bibitem[{{Lin} {et~al.}(2008){Lin}, {Yuan}, \& {Buta}}]{lin08}
{Lin}, L.-H., {Yuan}, C., \& {Buta}, R. 2008, \apj, 684, 1048

\bibitem[{{Lindblad} \& {Kristen}(1996)}]{lindblad96b}
{Lindblad}, P.~A.~B., \& {Kristen}, H. 1996, \aap, 313, 733

\bibitem[{{Lindblad} {et~al.}(1996){Lindblad}, {Lindblad}, \&
  {Athanassoula}}]{lindblad96}
{Lindblad}, P.~A.~B., {Lindblad}, P.~O., \& {Athanassoula}, E. 1996, \aap, 313,
  65

\bibitem[{{Lindt-Krieg} {et~al.}(2008){Lindt-Krieg}, {Eckart}, {Neri}, {Krips},
  {Pott}, {Garc{\'{\i}}a-Burillo}, \& {Combes}}]{lindt08}
{Lindt-Krieg}, E., {Eckart}, A., {Neri}, R., {et~al.} 2008, \aap, 479, 377

\bibitem[{{Maciejewski}(2004)}]{macie04}
{Maciejewski}, W. 2004, \mnras, 354, 892

\bibitem[{{Maoz} {et~al.}(1996){Maoz}, {Barth}, {Sternberg}, {Filippenko},
  {Ho}, {Macchetto}, {Rix}, \& {Schneider}}]{maoz96}
{Maoz}, D., {Barth}, A.~J., {Sternberg}, A., {et~al.} 1996, \aj, 111, 2248

\bibitem[{{Martin} \& {Kennicutt}(2001)}]{martin01}
{Martin}, C.~L., \& {Kennicutt}, Jr., R.~C. 2001, \apj, 555, 301

\bibitem[{{Nemmen} {et~al.}(2011){Nemmen}, {Storchi-Bergmann}, \&
  {Eracleous}}]{nemmen12}
{Nemmen}, R., {Storchi-Bergmann}, T., \& {Eracleous}, M. 2011, ArXiv e-prints

\bibitem[{{Ondrechen} {et~al.}(1989){Ondrechen}, {van der Hulst}, \&
  {Hummel}}]{ondrechen89}
{Ondrechen}, M.~P., {van der Hulst}, J.~M., \& {Hummel}, E. 1989, \apj, 342, 39

\bibitem[{{Patsis} \& {Athanassoula}(2000)}]{patsis00}
{Patsis}, P.~A., \& {Athanassoula}, E. 2000, \aap, 358, 45

\bibitem[{{Perez-Olea} \& {Colina}(1996)}]{perez96}
{Perez-Olea}, D.~E., \& {Colina}, L. 1996, \apj, 468, 191

\bibitem[{{Piner} {et~al.}(1995){Piner}, {Stone}, \& {Teuben}}]{piner95}
{Piner}, B.~G., {Stone}, J.~M., \& {Teuben}, P.~J. 1995, \apj, 449, 508

\bibitem[{{Quillen} {et~al.}(1995){Quillen}, {Frogel}, {Kuchinski}, \&
  {Terndrup}}]{quillen95}
{Quillen}, A.~C., {Frogel}, J.~A., {Kuchinski}, L.~E., \& {Terndrup}, D.~M.
  1995, \aj, 110, 156

\bibitem[{{Roberts}(1969)}]{roberts69}
{Roberts}, W.~W. 1969, \apj, 158, 123

\bibitem[{{Sakamoto} {et~al.}(1999){Sakamoto}, {Okumura}, {Ishizuki}, \&
  {Scoville}}]{sakamoto99}
{Sakamoto}, K., {Okumura}, S.~K., {Ishizuki}, S., \& {Scoville}, N.~Z. 1999,
  \apj, 525, 691

\bibitem[{{Sandage} \& {Tammann}(1981)}]{sandage81}
{Sandage}, A., \& {Tammann}, G.~A. 1981, {A revised Shapley-Ames Catalog of
  bright galaxies}

\bibitem[{{Sandstrom} {et~al.}(2010){Sandstrom}, {Krause}, {Linz},
  {Schinnerer}, {Dumas}, {Meidt}, {Rix}, {Sauvage}, {Walter}, {Kennicutt},
  {Calzetti}, {Appleton}, {Armus}, {Beir{\~a}o}, {Bolatto}, {Brandl},
  {Crocker}, {Croxall}, {Dale}, {Draine}, {Engelbracht}, {Gil de Paz},
  {Gordon}, {Groves}, {Hao}, {Helou}, {Hinz}, {Hunt}, {Johnson}, {Koda},
  {Leroy}, {Murphy}, {Rahman}, {Roussel}, {Skibba}, {Smith}, {Srinivasan},
  {Vigroux}, {Warren}, {Wilson}, {Wolfire}, \& {Zibetti}}]{sandstrom10}
{Sandstrom}, K., {Krause}, O., {Linz}, H., {et~al.} 2010, \aap, 518, L59

\bibitem[{{Schaye}(2004)}]{schaye04}
{Schaye}, J. 2004, \apj, 609, 667

\bibitem[{{Schwarz}(1984)}]{schwarz84}
{Schwarz}, M.~P. 1984, \mnras, 209, 93

\bibitem[{{Shlosman} {et~al.}(1989){Shlosman}, {Frank}, \&
  {Begelman}}]{shlosman89}
{Shlosman}, I., {Frank}, J., \& {Begelman}, M.~C. 1989, \nat, 338, 45

\bibitem[{{Skillman}(1987)}]{skillman87}
{Skillman}, E.~D. 1987, in NASA Conference Publication, Vol. 2466, NASA
  Conference Publication, ed. C.~J. {Lonsdale Persson}, 263--266

\bibitem[{{Storchi-Bergmann} {et~al.}(1993){Storchi-Bergmann}, {Baldwin}, \&
  {Wilson}}]{storchi93}
{Storchi-Bergmann}, T., {Baldwin}, J.~A., \& {Wilson}, A.~S. 1993, \apjl, 410,
  L11

\bibitem[{{Storchi-Bergmann} {et~al.}(1996){Storchi-Bergmann}, {Wilson}, \&
  {Baldwin}}]{storchi96}
{Storchi-Bergmann}, T., {Wilson}, A.~S., \& {Baldwin}, J.~A. 1996, \apj, 460,
  252

\bibitem[{{Telesco} {et~al.}(1993){Telesco}, {Dressel}, \&
  {Wolstencroft}}]{telesco93}
{Telesco}, C.~M., {Dressel}, L.~L., \& {Wolstencroft}, R.~D. 1993, \apj, 414,
  120

\bibitem[{{van der Hulst} {et~al.}(1987){van der Hulst}, {Skillman},
  {Kennicutt}, \& {Bothun}}]{hulst87}
{van der Hulst}, J.~M., {Skillman}, E.~D., {Kennicutt}, R.~C., \& {Bothun},
  G.~D. 1987, \aap, 177, 63

\bibitem[{{Wada}(2008)}]{wada08}
{Wada}, K. 2008, \apj, 675, 188

\bibitem[{{Wada} \& {Habe}(1992)}]{wada92}
{Wada}, K., \& {Habe}, A. 1992, \mnras, 258, 82

\bibitem[{{Wong} \& {Blitz}(2002)}]{wong02}
{Wong}, T., \& {Blitz}, L. 2002, \apj, 569, 157

\bibitem[{{Yang} {et~al.}(2007){Yang}, {Gruendl}, {Chu}, {Mac Low}, \&
  {Fukui}}]{yang07}
{Yang}, C.-C., {Gruendl}, R.~A., {Chu}, Y.-H., {Mac Low}, M.-M., \& {Fukui}, Y.
  2007, \apj, 671, 374

\bibitem[{{Yen} {et~al.}(2012){Yen}, {Taam}, {Yeh}, \& {Jea}}]{yen12}
{Yen}, C.-C., {Taam}, R.~E., {Yeh}, K.~H.-C., \& {Jea}, K.~C. 2012, Journal of
  Computational Physics, 231, 8246

\bibitem[{{Yuan} \& {Kuo}(1997)}]{yuankuo97}
{Yuan}, C., \& {Kuo}, C.-L. 1997, \apj, 486, 750

\bibitem[{{Yuan} \& {Kuo}(1998)}]{yuankuo98}
---. 1998, \apj, 497, 689

\bibitem[{{Yuan} \& {Yang}(2006)}]{yuan06}
{Yuan}, C., \& {Yang}, C.-C. 2006, \apj, 644, 180

\bibitem[{{Yuan} \& {Yen}(2005)}]{yuan05}
{Yuan}, C., \& {Yen}, D.~C.~C. 2005, Journal of Korean Astronomical Society,
  38, 197

\end{thebibliography}
\end{document}